\documentclass{aa501}

\usepackage{graphicx}
\begin{document}

\def\bh{{\rm BH}}


%

\title{Black hole mass and velocity dispersion of narrow line region in 
          active galactic nuclei and narrow line Seyfert 1 galaxies}
\author{Tinggui Wang \and Youjun Lu}

\offprints{Tinggui Wang (twang@ustc.edu.cn)} 

\institute{Center for Astrophysics, University of Science and Technology
              of China, Hefei, Anhui 230026, P.R. China \\
              National Astronomical Observatories, Chinese Academy of Sciences
             }

\date{Received 6 June 2001 / Accepted 12 July 2001 }

\authorrunning{T. Wang \& Y. Lu}
\titlerunning{Relation between the MBH mass and [OIII] width
                 in AGN}

\abstract{ 
   Controversy regarding whether Narrow Line Seyfert 1 galaxies (NLS1s) 
   follow the same black hole --- galactic bulge relation as normal galaxies
   has been raised by Mathur et al. (2001) and Ferrarese et al. (2001). In 
   this paper, the correlation between the black hole mass and the velocity 
   dispersion of the narrow line region, indicated by the [OIII] line 
   width for a heterogeneous sample of NLS1s, is examined. We show that 
   the non-virial component subtracted [OIII] width, which may well represent 
   the stellar velocity dispersion ($\sigma$), correlates with the estimated 
   black hole
   mass $M_\bh$, made using the empirical law for broad line region size
   from the reverberation mapping measurements. Considering measurement 
   errors, we find that the relation between $M_\bh$ and the [OIII] width
   in active galactic nuclei (both NLS1s in this paper and normal Seyferts 
   in Nelson 2000) is consistent with that defined in nearby hot galaxies 
   (ellipticals and spiral bulge) but with more scatter. It seems that there 
   is no clear difference in the relation between $M_\bh$ and $\sigma$ 
   (represented by the [OIII] width for AGN) in both NLS1s and normal broad
   line active galactic nuclei from the same relation defined by nearby 
   hot galaxies.
}
\maketitle

\section{Introduction}

Massive Black Holes (MBHs) are believed to exist in the centers of all active
and many or most normal galaxies. High-resolution observations of various 
kinematic tracers of the central gravitational potential have resulted in 
the detections of numerous MBHs in nearby galaxies over the past decade 
(e.g. \cite{kr95,fm00,geb00a,sar00,ho99}, and references therein). 
A relationship between the MBH mass and the mass of the spheroidal component
was suggested by Kormendy (1993) and later quantified by Kormendy \& 
Richstone (1995) and Magorrian et al. (1998). This correlation appears also 
in Seyfert 1 galaxies and QSOs, in which the MBH masses are measured either
using reverberation mapping techniques (\cite{wan99}) or using the
empirical relation between the Broad Line Region (BLR) size and nuclear 
luminosity (\cite{lao98}). Laor (1998) found that the MBH mass-to-bulge mass
ratio for a sample of PG QSOs is the same as that for nearby galaxies.
In contrast, Wandel (1999) obtained a substantially lower MBH mass-to-bulge
mass ratio for Seyfert 1 galaxies.  This could be due to an overestimation
of the bulge mass in Seyfert galaxies, e.g., a larger luminosity-to-mass 
ratio in the host galaxies, or an underestimation of central black
hole masses using the reverberation mapping method (\cite{md00,kro00}),
or an intrinsic difference in the MBH mass-to-bulge mass ratio for Seyfert 
galaxies and normal galaxies.

For nearby hot galaxies (ellipticals and spiral bulge), recent works by 
Gebhardt et al. (2000a) and 
Ferrarese \& Merritt (2000) have demonstrated that the mass of a MBH is tightly
correlated with the stellar velocity dispersion, which is obtained within
a large aperture extending to the galaxy effective radius and thus with
little influence of the MBH, with remarkably small scatter.  Note that Gebhardt
et al. (2000b) included also seven AGN, in which the MBH masses are obtained
by the reverberation mapping method, and they found that these objects 
follow the same correlation with small scatter. Ferrarese et al. (2001)
reached the same conclusion by making an accurate measurement of stellar
velocity dispersions for 6 Seyfert galaxies, for which the masses of MBHs
have been measured using reverberation mapping techniques.

The good correlation between the [OIII] width and the stellar velocity
dispersion (\cite{nw96}) indicates that the narrow-line width is primarily
virial in origin and the Narrow Line Region (NLR) kinematics are mainly
controlled by the gravitational potential of the galaxy bulge. For a sample
of 32 AGN and QSOs in which the MBH masses have been measured from
reverberation mapping, Nelson (2000) demonstrated a good relation between
the MBH mass and the bulge velocity dispersion derived from the [OIII] width,
which is consistent with the results of Gebhardt et al. (2000b) but with
somewhat larger scattering. This agreement can be taken as evidence in
support of the reverberation mapping method to measure the MBH masses in AGN.

The tight $M_\bh-\sigma$ relation supports the theoretical arguments
of a close link between the growth of MBHs and the galaxy or spheroidal
formation. Several theoretical scenarios have been proposed to explain
the $M_\bh-\sigma$ or $M_\bh-M_{\rm bulge}$ relation (e.g. 
\cite{sr98,fab99,ost00,hk00}). Silk \& Rees (1998) predicted 
$M_{\rm BH}\propto\sigma^5$, based on the back-reaction mechanism such that
the kinetic energy associated with the output wind from the central 
BH-accretion disk system will evacuate the fueling gas when it is
comparable to the bound energy of the gas in the bulge or host galaxy. Fabian
(1999) further incorporated the Silk--Rees scenario into an obscured growth
of MBHs model, and a consequent result is that most MBHs grow very
fast in an obscured phase before they clean the surrounding dust and cold gas
and appear as QSOs or AGN. This scenario can also explain both the
$M_{\rm BH}-\sigma$ and $M_{\rm BH}-M_{\rm bulge}$ relation.

It is of particular interest to investigate the time evolution (or 
accretion history) of MBHs and thus reveal the physical link between 
the bulge formation and the MBH growth. One approach is to measure 
the masses of MBHs and bulge properties in high redshift QSOs and AGN
and compare them with low redshift QSOs, AGN and nearby galaxies. Narrow
Line Seyfert 1 galaxies (NLS1s) are suggested to be due to accretion rates
close to the Eddington limit and have small BHs compared to normal Seyfert
1 galaxies at a given luminosity, and much evidence suggests that NLS1s 
might be normal Seyfert galaxies at an early stage of evolution 
(\cite[and references therein]{mat00}). If this is true, NLS1s could
be an ideal class of objects, together with normal Seyfert galaxies and
QSOs, to study the accretion history and growth of MBHs. Therefore, it is
also interesting to measure the masses of MBHs and bulge properties in 
NLS1s and compare them with those in Broad Line (BL) Seyfert 1 galaxies 
and nearby galaxies. Using the MBH mass estimated from spectra fitting by
an accretion disk model and the virial mass of the broad line region, Mathur
et al. (2001) found that NLS1s show systematically lower $M_\bh$ than 
BL AGN with the same bulge luminosity or [OIII] width of host galaxies. 
However, two NLS1s in the Ferrarese et al. (2001) sample follow the same 
relation as BL Seyfert 1 galaxies. The conflicting results in the literature 
suggest that further study is required. In the present paper, we investigate
the correlation between the MBH mass and the [OIII] line width for a large
sample of NLS1s (\cite{vvg}) and find that NLS1s consistently follow 
the well-known $M_\bh-\sigma$ relation defined in nearby galaxies.

\section{Data and Analysis}

A heterogeneous sample of 59 NLS1 galaxies were observed spectroscopically
by Veron-Cetty et al. (2001) (hereafter VVG) with a moderate resolution of
3.4\AA~. The measurement of the instrument-subtracted [OIII] and H$\beta$
width as well as the optical magnitude at B band are listed in table~1. 
VVG found that, in general, the [OIII] lines of those NLS1s have a 
relatively narrow Gaussian profile (with Full Width at Half Maximum, 
hereafter FWHM, of $\sim200-500$ km~s$^{-1}$) with often, in addition, a 
second broad blueshifted Gaussian component (with FWHM of $\sim 500-1800$ 
km~s$^{-1}$). The blueshifted Gaussian component is
proposed to be associated with those individual high-velocity clouds seen
in the spatially resolved NLR of some nearby Seyfert galaxies, which is 
outflowing rather than virial bounded (e.g. NGC~4151: \cite{kaiser}). The
width of the narrow component is then adopted as the velocity dispersion of
the virial NLR clouds if the line is fitted by a narrow component and a
blueshifted broad component.

\begin{table*}
  \caption{Black hole mass estimates and narrow line cloud
velocity dispersions. Col. 1: name, col. 2: B magnitude, col. 3: redshift, col.
4: the the Galactic hydrogen column density in units of 10$^{20}$ cm$^{-2}$,
col. 5: FWHM (in km s$^{-1}$) of the broad component of H$\beta$ line, col. 6:
FWHM (in km s$^{-1}$) of the broad component of [OIII] line, col. 7: the 
estimated size of broad line region using the empirical law (see \S\,2.1) and
col. 8: the estimated black hole mass in unit $10^7$M$_{\odot}$(see \S\,2.1).
~$^a$ the black hole mass has been measured in Kaspi et al. (2000) using
reverberation mapping techniques. }
\scriptsize
\begin{center}
\begin{tabular}{lccccccc} \hline \hline
Name & B & Z & NH & FWHM(H$\beta$) & FWHM([OIII]) & R$_{\rm BLR}$ & M$_{\rm BH}$ \\ \hline
 Mrk\,335$^a$       & 13.7 & 0.025 &  3.8 & $\cdots$ &  245 & $\cdots$ & 0.63     \\
 I\,ZW\,1           & 14.0 & 0.061 &  5.1 & $\cdots$ & 1040 & $\cdots$ & $\cdots$ \\ 
 Ton\,S\,180        & 14.4 & 0.062 &  1.5 & 1085 &  435 &  89.8 &   1.16 \\
 Mrk\,359           & 14.2 & 0.017 &  4.8 &  900 &  180 &  19.0 &   0.17 \\
 MS\,01442-0055     & 15.6 & 0.080 &  2.8 & 1100 &  240 &  63.7 &   0.85 \\
 Mrk\,1044          & 14.3 & 0.016 &  3.0 & 1010 &  335 &  15.0 &   0.17 \\
 HS\,0328+0528      & 16.7 & 0.046 &  8.9 & 1590 &  220 &  18.9 &   0.53 \\
 IRAS\,04312+40     & 15.2 & 0.020 & 34.5 &  860 &  380 &  52.2 &   0.42 \\
 IRAS\,04416+12     & 16.1 & 0.089 & 14.1 & 1470 &  650 &  92.8 &   2.20 \\
 IRAS\,04576+09     & 16.6 & 0.037 & 13.5 & 1210 &  380 &  18.5 &   0.30 \\
 IRAS\,05262+44     & 13.6 & 0.032 & 38.3 &  740 &  365 & 342.3 &   2.06 \\
 RXJ\,07527+261     & 17.0 & 0.082 &  5.1 & 1185 &  400 &  29.9 &   0.46 \\
 Mrk\,382           & 15.5 & 0.034 &  5.8 & 1280 &  155 &  23.0 &   0.41 \\
 Mrk\,705           & 14.9 & 0.028 &  4.0 & 1790 &  365 &  23.6 &   0.83 \\
 Mrk\,707           & 16.3 & 0.051 &  4.7 & 1295 &  315 &  23.2 &   0.43 \\
 Mrk\,124           & 15.3 & 0.056 &  1.3 & 1840 &  380 &  43.0 &   1.60 \\
 Mrk\,1239          & 14.4 & 0.019 &  4.1 & 1075 &  400 &  18.9 &   0.24 \\
 IRAS\,09571+84     & 17.0 & 0.092 &  3.9 & 1185 &  240 &  33.4 &   0.52 \\
 PG\,1011-040       & 15.5 & 0.058 &  4.5 & 1455 &  400 &  46.3 &   1.08 \\
 PG\,1016+336       & 15.9 & 0.024 &  1.6 & 1590 &  315 &   8.9 &   0.25 \\
 Mrk\,142           & 15.8 & 0.045 &  1.2 & 1370 &  260 &  22.7 &   0.47 \\
 KUG\,1031+398      & 15.6 & 0.042 &  1.4 &  935 &  315 &  23.6 &   0.23 \\
 RXJ\,10407+330     & 16.5 & 0.081 &  2.2 & 1985 &  460 &  35.3 &   1.53 \\
 Mrk\,734           & 14.6 & 0.049 &  2.7 & 1825 &  180 &  59.6 &   2.18 \\
 Mrk\,739E          & 14.1 & 0.030 &  2.2 & 1615 &  380 &  39.9 &   1.14 \\
 MCG\,06.26.012     & 15.4 & 0.032 &  1.9 & 1145 &  220 &  18.6 &   0.27 \\
 Mrk\,42            & 15.4 & 0.024 &  1.9 &  865 &  220 &  12.4 &   0.10 \\
 NGC\,4051$^a$      & 12.9 & 0.002 &  1.3 & 1120 &  200 &   1.8 &   0.13 \\
 PG\,1211+143$^a$   & 14.6 & 0.085 &  2.8 & 1975 &  410 & 132.6 &   4.05 \\
 Mrk\,766           & 13.6 & 0.012 &  1.8 & 1630 &  220 &  14.8 &   0.43 \\
 MS\,12170+0700     & 16.3 & 0.080 &  2.2 & 1765 &  365 &  39.4 &   1.35 \\
 MS\,12235+2522     & 16.3 & 0.067 &  1.8 &  800 &  240 &  29.9 &   0.21 \\
 IC\,3599           & 15.6 & 0.021 &  1.4 & $\cdots$ &  280 &   8.8 & $\cdots$ \\
 PG\,1244+026       & 16.1 & 0.048 &  1.9 &  740 &  330 &  21.2 &   0.13 \\
 NGC\,4748          & 14.0 & 0.014 &  3.6 & 1565 &  295 &  15.5 &   0.42 \\
 Mrk\,783           & 15.6 & 0.067 &  2.0 & 1655 &  430 &  47.4 &   1.43 \\
 R\,14.01           & 14.6 & 0.042 &  7.6 & 1605 &  430 &  60.5 &   1.71 \\
 Mrk\,69            & 15.9 & 0.076 &  1.1 & 1925 &  315 &  44.9 &   1.83 \\
 2E\,1346+2646      & 16.5 & 0.059 &  1.1 & $\cdots$ &  180 &  21.2 & $\cdots$ \\
 PG\,1404+226       & 15.8 & 0.098 &  2.0 & 1120 &  950 &  72.4 &   1.00 \\
 Mrk\,684           & 14.7 & 0.046 &  1.5 & 1150 & 1290 &  48.2 &   0.70 \\
 Mrk\,478           & 14.6 & 0.077 &  1.0 & 1270 &  365 & 105.3 &   1.87 \\
 PG\,1448+273       & 15.0 & 0.065 &  2.7 & 1050 &  155 &  69.1 &   0.84 \\
 MS\,15198-0633     & 14.9 & 0.084 & 12.4 & $\cdots$ & $\cdots$ & 170.4 & $\cdots$ \\
 Mrk\,486           & 14.8 & 0.038 &  1.8 & 1680 &  400 &  34.9 &   1.08 \\
IRAS\,15462-0450    & 16.4 & 0.100 & 12.5 & 1615 & 1600 &  83.9 &   2.40 \\
 Mrk\,493           & 15.1 & 0.031 &  2.0 &  740 &  315 &  21.7 &   0.13 \\
 EXO\,16524+393     & 16.7 & 0.069 &  1.7 & 1355 &  400 &  24.0 &   0.48 \\
 B\,31702+457       & 15.1 & 0.060 &  2.2 &  975 &  295 &  56.4 &   0.59 \\
 RXJ\,17450+480     & 15.9 & 0.054 &  3.1 & 1355 &  400 &  30.2 &   0.61 \\
 Kaz\,163           & 15.0 & 0.063 &  4.4 & 1875 &  480 &  71.7 &   2.77 \\
 Mrk\,507           & 15.4 & 0.053 &  4.3 & 1565 & 1025 &  43.0 &   1.16 \\
 HS\,1817+5342      & 15.2 & 0.080 &  4.9 & 1615 &  570 &  91.2 &   2.61 \\
 HS\,1831+5338      & 15.9 & 0.039 &  4.9 & 1555 &  240 &  20.7 &   0.55 \\
 Mrk\,896           & 14.6 & 0.027 &  4.0 & 1135 &  315 &  27.1 &   0.38 \\
 MS\,22102+1827     & 16.7 & 0.079 &  6.2 &  690 &  890 &  36.2 &   0.19 \\
 Akn\,564           & 14.2 & 0.025 &  6.4 &  865 &  220 &  35.4 &   0.29 \\
 HS\,2247+1044      & 15.8 & 0.083 &  6.2 & 1790 &  710 &  69.6 &   2.45 \\
 Kaz\,320           & 16.8 & 0.034 &  4.9 & 1470 &  260 &   9.5 &   0.23 \\
\hline

\end{tabular}
\end{center}
\end{table*}

\subsection{Estimation of black hole masses}

The size of the broad emission line region (BLR) can be estimated
by the empirical relationship between the size and the
monochromatic continuum luminosity at 5100\AA~ (\cite{kas00}):
\begin{equation}
R_{\rm BLR} = 32.9 \left(\frac{\lambda L_{\lambda}(5100{\rm \AA})}
{10^{44}{\rm erg\cdot s^{-1}}} \right)^{0.7} {\rm lt~day}~,
\label{eq:blr}
\end{equation}
where $\lambda L_{\lambda}$ is estimated from the B-magnitude by adopting
an average optical spectral index of $-0.5$ and accounting for
Galactic reddening and K-correction (H$_0$=75km~s$^{-1}$~Mpc$^{-1}$, 
q$_0$=0.5). Assuming that the BLR is virialized, the MBH mass can be estimated
by $M_{\rm BH} = R_{\rm BLR} V^2 G^{-1}$, where $G$ is the gravitational
constant, $V$ can be estimated from the emission line width,
$V=\sqrt{3}/2{\rm FWHM}$, by assuming BLR clouds in random orbit motion.
The estimated BH masses are also listed in Table~1.

There are some uncertainties in the estimation of the MBH mass. First,
a typical error of 0.2 mag in the B magnitude given in VVG would
introduce an uncertainty of above 0.05 dex in MBH mass. Second, the
continua are likely to be variable, but generally this variation 
is not larger than a factor of 2 for most AGN (cf. \cite{kas00}), which
may introduce an uncertainty of 0.15 dex in the estimation of MBH mass. 
Third, using the empirical law of equation~\ref{eq:blr} to estimate the 
BLR size, the uncertainties are generally not much larger than a factor
of 2 for those NLS1s in VVG sample (see \cite{kas00}) with
$\lambda L_{\lambda}(5100{\rm \AA})$ range from $10^{43}$ to $10^{45}$ 
erg~s$^{-1}$, if those NLS1s do follow this empirical relation. Finally,
a significant fraction of optical light may come from host galaxies. To
make a quantitative estimation of this effect in a NLS1, we notice that
$L_{\rm AGN}/L_{\rm bulge}=L_{\rm AGN}/L_{\rm Edd}\times L_{\rm Edd}/M_\bh
\times M_{\rm bulge}/L_{\rm bulge}\times M_\bh/M_{\rm bulge}$, and 
$L_{\rm Edd}/M_{\rm BH}\sim 3\times 10^4 L_{\sun}/M_{\sun}$. For NLS1,
the typical value of $L_{\rm AGN}/L_{\rm Edd}$ should be around 0.5 
(Puchnarewicz et al. 2001);
the typical bulge mass to light ratio may be similar to nearby hot
galaxies with $M_{\rm bulge}/L_{\rm bulge}\sim 10 M_{\sun}/L_{\sun}$;
and the MBH mass to bulge mass ratio may be similar to (or less than) 
nearby galaxies with $M_\bh/M_{\rm bulge}$ of about 0.0015 --- 0.003 
(by an order of magnitude) (Merritt \& Ferrarese 2001b;
Gebhardt et al. 2000a; Mathur et al. 2001). The fraction of light at 
the optical band $L_{\rm opt}/L_{\rm bol}$ is $\sim 0.1$ for AGN and 
$\sim 1.0$ for bulge. Adopting those values, one obtains 
$L_{\rm opt,AGN}/L_{\rm opt,bulge}\ga 1-10$, which suggests that  
the stellar contribution to the measured optical luminosity should
be much less (or less) than that from the nuclear emission. 
Thus, the uncertainty in the mass estimation is small in comparison with
the intrinsic scatter in the mass of the sample. Combining all those 
uncertainties, the estimation of 
the MBH mass would typically have an uncertainty of about 0.5 dex.

\subsection{Estimation of the bulge velocity dispersion}
\label{sec:oiii}

The [OIII] width can be converted to the stellar velocity dispersion by
$\sigma={\rm FWHM}_{\rm [OIII]}/2.35$ (\cite{nw95}). Nelson (2000) has 
shown that the reverberation mapping measured MBH mass in AGN, for which 
the bulge velocity dispersion is derived from the [OIII] width, is in 
good agreement with the $M_{\rm BH}-\sigma$ relation defined by nearby 
hot galaxies, which may support the assertion 
that the narrow line [OIII] width serves
as a good representation of the bulge velocity dispersion. The $\sigma$
derived this way is systematically lower than the stellar velocity 
dispersion from the absorption line width by 0.1 dex, while the mean 
deviation to the best fit line is 0.13 dex (\cite{nw95}). 
This systematic difference will not affect the 
following statistical analysis significantly since it is much smaller 
than the intrinsic scatter in the [OIII] line width measurements.

The [OIII] width could be significantly over-estimated from the spectra with
poor resolution (Veilleux 1991, hereafter V91). According to Figure~3 in
V91, this overestimation could be as large as a factor of 1.2---1.5 if the
spectral resolution is close to the intrinsic [OIII] width of the object. 
Note that three objects in VVG, 
Mrk~359, NGC~4051, Mrk~766, were also observed by 
Veilleux (1991) at a resolution of 10 km~s$^{-1}$, and the [OIII] line width
(the width measured by VVG, the V91 width to the VVG width ratio) are 113 
(180, 1.59), 162 (200, 1.23) and 180 (220, 1.22) km~s$^{-1}$, respectively. 
These values clearly support that the [OIII] line width is overestimated by
a factor of 1.2---1.5 for those objects with intrinsic widths close to or 
less than the spectral resolution, i.e. 204 km~s$^{-1}$. As discussed by 
Whittle (1985), the detailed amount of the deviation is also sensitive 
to the line profile. This would suggest that the measured line width of less
than 300 km~s$^{-1}$ may be overestimated by such a factor. In the following
analysis, we will keep in mind this uncertainty, and discuss its
consequences wherever appropriate. Note also that the [OIII] widths for most
objects in the Nelson (2000) sample were measured from the spectra with high 
resolution, $<2$\AA, corresponding to $<120$ km~s$^{-1}$, which may not
suffer from the overestimation due to spectral resolution, since measured
[OIII] line widths are much larger than the spectral resolution. 

\subsection{$M_{\rm BH}$ to the bulge velocity dispersion relation}

\begin{figure*}
   \centering
   \includegraphics[width=17cm]{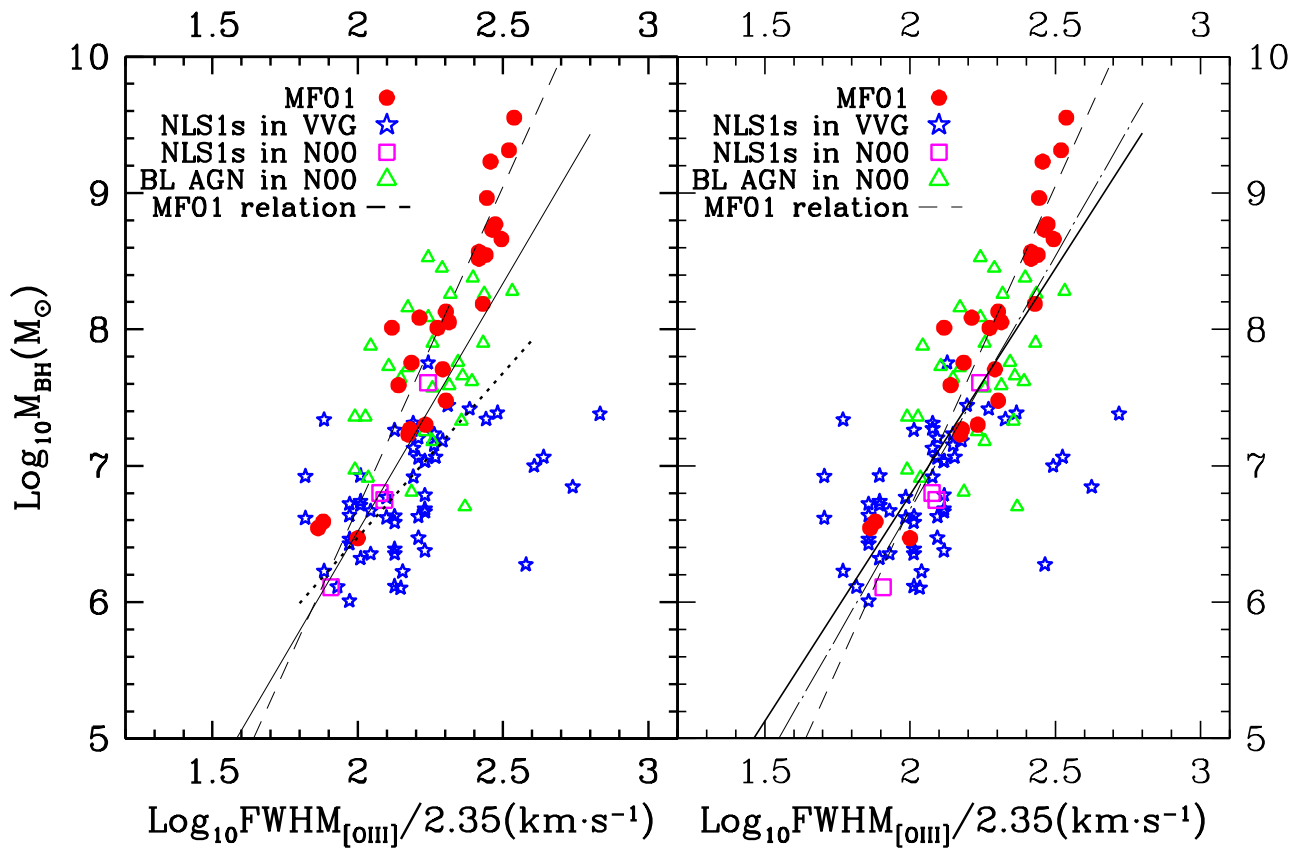}
   \caption{The estimated mass of MBHs versus the stellar
            velocity dispersion derived from the [OIII] line width:
            NLS1s from Veron-Cetty et al. (2001) (VVG) are shown as open
            stars. For comparison, NL AGN and BL AGN from Nelson
            (2000) (N00) are plotted as open squares and open triangles,
            respectively. The solid circles represent the nearby
            hot galaxies from Merritt \& Ferrarese (2001a) (MF01). The dashed
            line is the best fitted line for nearby hot galaxies
            (MF01). The [OIII] width is corrected
            for the possible overestimation due to the low spectral
            resolution by a factor of 1.3 in the right panel, but not
            in the left panel. In the left panel, the dotted (solid)
            line is the best fitted line for NL AGN (NL + BL AGN,
            78 objects in total by excluding
            those six objects which deviate from the others and have the
            largest [OIII] width, see also \S 2.3);
            in the right panel, the solid (dotted long-dashed) line
            is the best fitted line for the 78 NL + BL AGN ( 75 NL + BL
            AGN by excluding those
            three objects which deviate from the main trend but have the
            smallest [OIII] width, see also \S2.3).}
   \label{fig:msigma_mf}
\end{figure*}

\begin{figure*}
   \centering
   \includegraphics[width=17cm]{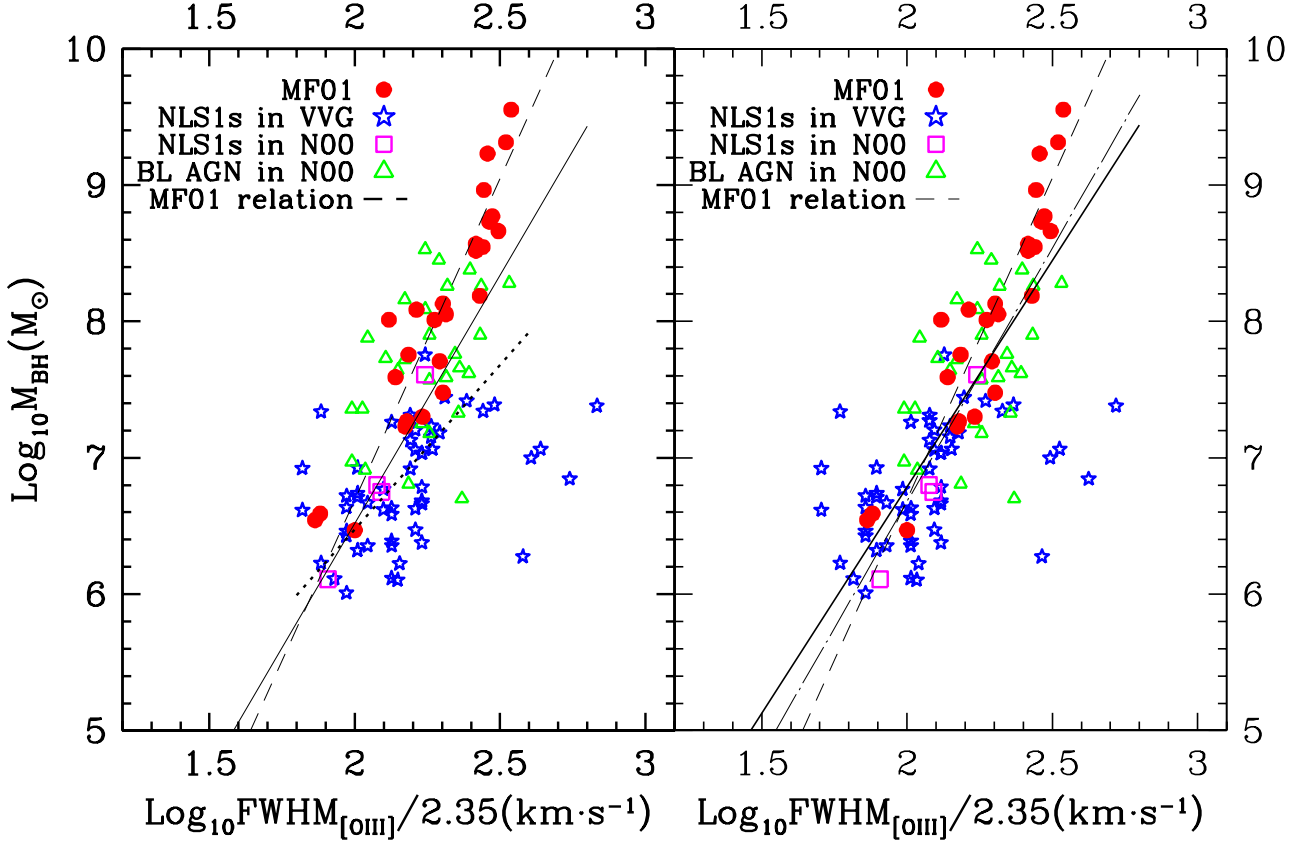}
   \caption{Legend as Fig.~\ref{fig:msigma_mf}, but adopting the galaxies
            from Gebhardt et al. (2000a) (G00) for comparison. It
            is obviously that the best fit slope of AGN is 
            consistent with the one derived by G00. }
   \label{fig:msigma_geb}
\end{figure*}

The relationship between the estimated MBH mass $M_{\rm BH}$ and the bulge
velocity dispersion represented by the [OIII] width is shown in the left
panel of Figures~\ref{fig:msigma_mf} and \ref{fig:msigma_geb} for NLS1s 
in VVG, along with the same 
relationship for those BL AGN in Nelson (2000). Note that four objects,
NGC~4051, Mrk~335, PG~1211+143 and Mrk~110, which have reverberation
mapping measured MBH masses and high resolution ($R\ga1500$) [OIII] line
widths, in the Nelson (2000) sample are NL Seyfert 1 galaxies or QSOs. The
former three objects are also included in VVG. The high quality data 
in Nelson (2000) are adopted for these three objects instead of the estimated
ones from VVG. The effectiveness of using the empirical law to estimate
the mass of MBHs in NLS1s may also be supported by the fact that the other NL
objects, whose masses are estimated from the empirical law, follow the trend
of these four NL objects in Figures~\ref{fig:msigma_mf} and 
\ref{fig:msigma_geb}. 
In the present paper, the stellar velocity dispersion derived from the
[OIII] width is likely to represent the central velocity dispersion.
For comparison, the nearby hot galaxies from Merritt \& Ferrarese (2001a), 
for which values of $M_{\rm BH}$ and $\sigma$ (the central stellar velocity
dispersion) are measured from dynamical modeling of {\it HST} data, are 
therefore also plotted in Figure~\ref{fig:msigma_mf}.
Since there are different views on the slope of the $M-\sigma$ relation defined
in nearby galaxies (see Merritt \& Ferrarese 2001a; Gebhardt et al. 2000a), 
a similar figure is also plotted (see Fig.~\ref{fig:msigma_geb}) for comparison
with the galaxies from Gebhardt et al. (2000a), in which the brightness
weighted stellar 
velocity dispersions within the effective radius of galaxies are adopted.

It is clearly shown in Figures~1 and 2 that five NL objects, Mrk~507, Mrk~684, 
IRAS~15462-0450, MS~22102+1827, and PG~1404+226, having large [OIII] 
widths, deviate from other NL objects. Note that Mrk~507, Mrk~684,
IRAS~15462-0450 are the three objects for which a narrow HII region
contribution has been subtracted in the VVG sample. For spectra with a 
resolution of 3.4\AA~ used by VVG, the HII component may not be reliably 
separated from a narrow component of NLR with width 200---500
km\,s$^{-1}$ if an additional broad wing is present. Thus the widths of NLR
in these three objects are likely to be significantly over-estimated.
Only poor quality [OIII] profiles are available for MS~22102+1827 and 
PG~1404+226, and VVG mentioned that broad blueshifted [OIII] profiles 
should not be overlooked in PG 1404+226. These five objects will be 
excluded in the following statistic analysis. 
The galactic bulge mass of IC~4329A, which clearly deviates from other
AGN in the Figure~1 in Nelson (2000), is one of the smallest in the Wandel
(1999) sample of about $10^{10.6}$ solar mass. However, the bulge velocity
dispersion derived from the [OIII] width of IC~4329A is one of the largest.
The small bulge mass but large bulge velocity in this object compared with
others in the Wandel (1999) sample is in contradiction with the Faber-Jackson 
relation. The
high resolution radio map of IC~4329A consists of a compact core and with
extended component to several kpc (Unger et al. 1987). If the extended
component is the radio jet, then the large [OIII] width can be due to the
non-virial component (Nelson \& Whittle 1996). This object will also be
removed from the sample in the following analysis.

Considering both the NL Seyfert 1 galaxies and NL QSOs in VVG and BL
AGN in the Nelson (2000) sample, a Spearman rank correlation tests gives 
a strong correlation between $M_{\rm BH}$ and $\sigma$ for 78 AGN with 
a correlation coefficient of $R_{\rm s}=0.613$ corresponding to 
a probability of $P_{\rm s}=2.4\times 10^{-9}$ that the correlation is 
caused by a random factor, which can be fitted by a line with a slope of 
$3.64\pm0.21$ using an ordinary least-squares (OLS) bisector\footnote{
One should be cautious, as the use of OLS can be very misleading: linear
fits not accounting for errors are known to underestimated the true slope
of the relation. Merritt \& Ferrarese (2001a), for example, have argued that
the shallower slope found for the M-$\sigma$ relation by Gebhardt et al. (2000)
is due mostly to a bias introduced by neglecting the observational errors.
In our sample, however, it is hard to assign errorbars to the BH masses and
velocity dispersions for galaxies.}
(\cite{isobe}) (represented by the solid line in the left panel of 
Figs.~\ref{fig:msigma_mf} and ~\ref{fig:msigma_geb}). This slope agrees well
with the one defined in nearby hot galaxies derived by Gebhardt et al. (200a),
but deviates from the one derived by Merritt \& Ferrarese (2001a),
and the MBH mass in AGN seems smaller than the one in nearby hot galaxies 
by 0.5~dex. If we only consider NL Seyfert 1 galaxies
and NL QSOs (51 objects), the correlation is also moderately significant
with $R_{\rm s}=0.553$ ($P_s=2.6\times 10^{-5}$), which can be fitted by
a line with a slope of $2.70\pm0.28$ using the OLS bisector (represented by 
the dotted line in the left panel of Figs.~\ref{fig:msigma_mf} and 
\ref{fig:msigma_geb}). Compared with
the $M_\bh-\sigma$ relation defined by nearby hot galaxies (Gebhardt et al. 
2000a), we find the MBH mass in NLS1s is smaller than that in nearby hot 
galaxies by $\sim 0.5$~dex. However, we may not be able to draw a conclusion
that the MBHs in NLS1s (or AGN) are systematically smaller than that in nearby
hot galaxies at a given bulge velocity dispersion if the uncertainties in
the estimation of MBH mass (about 0.5 dex) and possible overestimation of 
the [OIII] width (see following paragraph) are considered.

As discussed in \S~\ref{sec:oiii}, the low spectral resolution could
introduce an overestimation of the [OIII] width, probably by a factor
of 1.2---1.5. The relationship between $M_{\rm BH}$ and $\sigma$ are
re-plotted in the right panel of Figure~\ref{fig:msigma_geb} by correcting
this overestimation of a moderate factor 1.3 for NLS1s in VVG. Now, the
correlation between $M_{\rm BH}$ and $\sigma$ becomes very strong with
a coefficient $R_{s}=0.730$ ($P_s=3.8\times 10^{-14}$) for the combined
sample (78 AGN), and can be fitted (using the OLS bisector) by
\begin{equation}
M_{\rm BH} = 10^{(7.78\pm0.005)}M_{\odot}\left(
\frac{{\rm FWHM}_{\rm [OIII]}/2.35}{200{\rm km\cdot {\rm s}^{-1}}}
\right)^{3.32\pm0.38},
\end{equation}
as shown by the solid line in the right panel of Figures~\ref{fig:msigma_mf}
and \ref{fig:msigma_geb}. As seen in Figures~\ref{fig:msigma_mf} and 
\ref{fig:msigma_geb}, three objects,
Mrk~382, Mrk~734 and PG~1448+273, which have the smallest [OIII] widths,
deviate from the other NL objects. The reason that they have somewhat
larger MBHs than others is not known. However, if excluding them, the 
correlation becomes even stronger with $R_{s}=0.762$ 
($P_s=2.0\times 10^{-15}$), which can be fitted (using OLS bisector) by 
\begin{equation}
M_{\rm BH} = 10^{(7.81\pm0.006)}M_{\odot}\left(
\frac{{\rm FWHM}_{\rm [OIII]}/2.35}{200{\rm km\cdot {\rm s}^{-1}}}
\right)^{3.70\pm0.51},
\end{equation}
as shown by the dotted-long dashed line in the right panel of 
Figures~\ref{fig:msigma_mf} and \ref{fig:msigma_geb}. 
The scatter is large for this relation, which should be due to large
uncertainties in both variables. Whether we exclude Mrk~382, Mrk~734 and 
PG~1448+273 or not, the $M_\bh-\sigma$ relation for AGN (both NL and BL
AGN) is consistent with that defined in nearby hot galaxies. It seems
also that the slope of the fit agrees with the one found by Gebhardt et
al. (2000a) for nearby hot galaxies and Nelson (2000) for AGN (see Fig.~2),
but is different to the one derived by Merritt \& Ferrarese (2001a) of 4.72 
(see Fig.~1); more conclusive result
needs precise measurement of both the MBH mass and bulge velocity
dispersion. Again, the difference from the $M_\bh-\sigma$ relation
(Gebhardt et al.  2000a) in Log$_{10}$$M_\bh$ or Log$_{10}$$\sigma$ is
about -0.5
or 0.1. The consistency of the MBH mass in NL Seyfert 1 galaxies and NL
QSOs with the bulge velocity dispersion supports the result of Ferrarese
et al. (2001) that the two NLS1s with measured MBH masses (by the reverberation
mapping method) and bulge velocity dispersions (from stellar absorption
lines) are consistent with $M_{\rm BH}-\sigma$ relation defined by nearby
hot galaxies.

\section{Discussion and Conclusion}

In this paper, we find that there is no clear difference in the relation
between $M_{\rm BH}$ and $\sigma$ (the bulge velocity dispersion is
represented by the [OIII] width for AGN) for both NL and BL AGN from
the same relation defined by nearby hot galaxies (Gebhardt
et al. 2000a; Ferrarese \& Merritt 2000). Furthermore, the MBH masses
and bulge velocity dispersions of NLS1s are thought to be consistent
with the $M_{\rm BH}-\sigma$ relation for other galaxies if we consider the
overestimation in the [OIII] line width. This consistency suggests that
NLS1s have small MBHs compared with BL AGN with similar non-thermal
luminosity simply due to their host galaxies having small bulges compared 
to that of BL AGN. This may support one of the competing model of
NLS1s, i.e., the low-mass/high accretion rate interpretation.

A simple evolutionary scenario proposed by Mathur (2000) is that NLS1s are
likely to represent a crucial early and more obscured phase in the evolution
of active galaxies based on the observational properties of NLS1s, such
as super-solar metallicities and are unusually luminous in the far-infrared 
band etc. This evolutionary view has also been frequently suggested by other
authors (\cite{law00}). Although it is tentative that the $M_{\rm BH}$ and
$\sigma$ of NLS1s is consistent with the $M_{\rm BH}-\sigma$ relation 
(Gebhardt et al. 2000a, Ferrarese \& Merritt 2000), we cannot rule out the 
possibility that the MBHs in NLS1s are smaller than those in BL AGN or nearby
hot galaxies at a given bulge velocity dispersion by a factor of several
(say, 3), which means that Mathur's scenario cannot simply be ruled
out by our results. However, the claim of Mathur et al. (2001) that NLS1s
have a significantly smaller MBH to bulge velocity dispersion ratio,
which may be caused by some non-virial component in their [OIII] line
width measurements, is discredited by our results.   

Now we have more confidence in applying the reverberation mapping method to 
measure the masses of MBHs in AGN, since the MBH masses from reverberation
mapping are consistent with the $M_{\rm BH}-\sigma$ relation recently 
discovered for local galaxies (\cite{geb00b,fer01,ne00}). Krolik (2000)
pointed out, however, this consistence could be due to fortuitous mutual 
canceling of the systematic errors$-$including overestimation of the MBH 
mass by a fixed ratio by interpreting the emission line kinematics as
gravitationally bound and underestimating the mass for planar-like BLR
cloud distribution. If the narrowness of the permitted line width
of NLS1s is due to a planar-like BLR viewed nearly ``pole-on''
(the ``orientation model'': Osterbrock \& Pogge 1985; Goldrich
1989; Puchnarewicz et al. 1992), the estimated MBH masses would be
systematically smaller than the real one. Since the orientation
is a random effect, we would expect that the estimated MBH masses
in BL AGN are systematically larger than those in NLS1 by a
similar factor of 10, considering of NLS1s broad line width
are around $1000$~km~s$^{-1}$ while BL objects are typically about
$3000-5000$~km~s$^{-1}$ (which means that the velocity of broad-line-emitting
clouds would be underestimated by a factor of about 3 if both NLS1s
and BL objects have a similar central engine and a flat broad line geometry),
at a given stellar velocity dispersion. As we can see in 
Figure~\ref{fig:msigma_mf} and \ref{fig:msigma_geb}, 
masses of NL objects at a given bulge velocity dispersion are consistent
with the trend of BL objects, which suggests that at
least not all NLS1s can be regarded as ``orientation'' dependent.

It is generally believed that the activity in galactic nuclei
is closely linked with the galaxy and bulge formation. Silk \&
Rees (1998) proposed that the powerful wind from the central
engine can blow away the cold gas from the galaxy and terminate
the accretion process when the output kinetic energy is comparable
with the bound energy of the total gas in the galaxy. This results
in a relation of the MBH mass to the stellar velocity dispersion of
the form $M_{\rm BH}\propto \sigma^5$. The typical duration of
the bright QSOs phase is required to be only about few $10^7$ yr from
fitting the optical QSOs luminosity function by the mass function
of dark matter halos predicted by standard hierarchical
cosmogonies (\cite{hnr98}). It suggests that
MBHs may grow at an accretion rate far above the Eddington
rate before this brief optical bright phase and/or at a very low
accretion rate via advection-dominated accretion flows lasting a
Hubble time after this phase.
Fabian (1999) further incorporated the Silk-Rees scenario in a
model of obscured growth of MBHs. In his model, a MBH in the center
of a galaxy accretes the surrounding material and emits a QSO/AGN-like
spectrum which is absorbed by surrounding gas and dust. The
wind from the central engine exerts a force on the gas and pushes it
outwards. The central engine emerges when the Thomson depth in the
$M_{\rm BH}-\sigma$ relation for bright AGN or NL objects is similar to
that of the galaxies.
Our result of a consistent $M_{\rm BH}-\sigma$ relation in NLS1s favors
the model of Fabian (1999).

Many works have been done on the relation between $M_{\rm BH}$ and
$M_{\rm bulge}$. There is still controversy about whether Seyfert
galaxies have a small $M_{\rm BH}$ to $M_{\rm bugle}$ ratio compared
with local galaxies or not (Wandel 1999, McLure \& Dunlop 2000).
Czerny et al. (2000) claimed that NLS1s, at least, have a smaller 
$M_{\rm BH}$ to $M_{\rm bugle}$ ratio, which could be due to nuclear star
burst (or stellar formation and evolution) in NLS1s leading to a small
mass to light ratio of bulges. 

In the present paper, there are some caveats for both the estimation 
of the MBH mass and
the bulge velocity. First, the empirical $R_{\rm BLR}$-$L$ relation is not
fully tested for NLS1s. This relation is derived from a moderate-size sample
of AGN, composed mainly of BL AGN (\cite{kas00}). Three of the four NLS1s in
this sample closely follow the $R_{\rm BLR}$-$L$ relation. The lowest luminosity
object, NGC 4051, shows a larger size of BLR than this empirical relation 
predicted. There is, at least, no obvious evidence against this empirical 
relation, although further confirmation is needed. Second, Nelson \& 
Whittle (1996) identified two cases in which the [OIII] width can be 
significantly larger than the bulge velocity dispersion, i.e., presenting 
kpc linear radio sources or displaying distorting morphology. Though 
lacking in systematic study, NLS1s tend to possess similar radio properties 
to average radio-quiet Seyfert galaxies (\cite{uag95}). Zheng et al. 
(1999) found that NLS1s in their sample are morphology relaxed. We also 
note that two NL objects included in the sample of Nelson \& Whittle (1995) 
do not show systematic deviation. Also, the non-virial component of [OIII] 
lines has been subtracted for VVG objects. Therefore, our results should 
not be affected by the possible linear radio source in some objects.

\begin{acknowledgements}
   We thank an anonymous referee for helpful comments and suggestions.
   TW thanks the financial support from Chinese NSF through grant
   NSF-19925313 and from Ministry of Science and Technology. YL acknowledges
   the hospitality of the Department of Astrophysical Sciences, Princeton 
   University.
\end{acknowledgements}


\begin{thebibliography}{}
   \bibitem[Blandford 1999]{bla99}Blandford, R. D., 1999, in ``Origin and
            Evolution of Massive Black Holes in Galactic Nuclei'', ed. Merritt,
            Valluri \& Sellwood, 1999, p87
   \bibitem[Czerny et al. 2000]{czn00}Czerny, B., Nikolajuk, M., Piasecki, M.,
           \& Kuraszkiewicz, J., 2001, MNRAS, 325, 865
   \bibitem[Fabian 1999]{fab99}Fabian, A. C., 1999, MNRAS, 308, L39
   \bibitem[Ferrarese \& Merritt 2000]{fm00}Ferrarese, L., \& Merritt, D.
           2000, ApJL, 539, L9
   \bibitem[Ferrarese et al. 2001]{fer01} Ferrarese, L., Pogge, R. W.,
           Peterson, B. M., Merritt, D., Wandel, A., \& Joseph, C. L., 2001,
           ApJL, 555, L79
   \bibitem[Gebhardt et al. 2000a]{geb00a}Gebhardt, K., et al. 2000a,
           ApJL, 539, L13
   \bibitem[Gebhardt et al. 2000b]{geb00b}Gebhardt, K., et al. 2000b,
           ApJL, 543, L5
   \bibitem[Goldrich 1989]{gold89}Goldrich, R. W., 1989, ApJ, 342, 224
   \bibitem[Haehnelt \& Kauffmann 2000]{hk00}Haehnelt,M. G., Kauffmann, G., 
           2000, MNRAS, 318, L35
   \bibitem[Haehnelt et al. 1998]{hnr98}Haehnelt, M. G.,
           Natarajan, P. \& Rees, M. J., 1998, MNRAS, 300, 817
   \bibitem[Ho 1999]{ho99}Ho, L., 1999, in "Observational Evidence for the 
           Black Holes in the Universe", ed. S. K. Chakrabarti 
           (Dordrecht:Kluwer), p157
   \bibitem[Kaiser et al. 2000]{kaiser} Kaiser, M. E., Bradley, L. D. II,
           Hutchings, J. B., Crenshaw, D. M., Gull, T. R., Kraemer, S. B.,
           Nelson, C. H., Ruiz, J., \& Weistrop, D., 2000, ApJ, 528, 260
   \bibitem[Kaspi et al. 2000]{kas00}Kaspi, S., Smith, P. S., Netzer, H.,
           Maoz, D., Jannuzi, B. T., \& Giveon, U., 2000, ApJ, 533, 631
   \bibitem[Isobe et al. 1990]{isobe} Isobe, T., Feigelson, E. D., 
           Akritas, M. G., \& Babu, G. J., 1990, ApJ, 364, 104
   \bibitem[Kauffmann \& Haehnelt 2000]{kh00}Kauffmann, G., Haehnelt, M., 
           2000, MNRAS, 311, 576
   \bibitem[Kormendy 1993]{kor93}Kromendy, J., 1993, in The Nearest 
           Active Galaxies, eds. J. Beckman, L. Colina, \& H. Netzer 
           (Madrid: CSIC), 197
   \bibitem[Kormendy \& Richstone 1995]{kr95}Kormendy, J., \& Richstone, D.,
           1995, ARA\&A, 33, 581
   \bibitem[Krolik 2000]{kro00}Krolik, J., 2001, ApJ, 551, 72
   \bibitem[Laor 1998]{lao98}Laor, A., 1998, ApJL, 505, L83
   \bibitem[Law-Green et al. 2000]{law00}Law-Green, J. D. B., Hirst, P., 
           O'Brien, P. T., Ward, M., \& Boisson, C., 2000, MNRAS submitted
   \bibitem[Magorrian et al. 1998]{mag98}Magorrian, J., et al. 
           1998, AJ, 115, 2285
   \bibitem[Mathur 2000]{mat00}Mathur, S., 2000, MNRAS, 314, L17
   \bibitem[Mathur et al. 2001]{mat01}Mathur, S., Kuraszkiewicz, J., \&
           Czerny, B. 2001, New Astron., 6, 321
   \bibitem[McLure \& Dunlop 2000]{md00}McLure, R. J., \& Dunlop, J. S.,
           2000, MNRAS, in press (astro-ph/0009406)
   \bibitem[Merritt \& Ferrarese 2001a]{mf01a} Merritt D, \& Ferrarese L.,
           2001a, ApJ, 547, 140
   \bibitem[Merritt \& Ferrarese 2001b]{mf01m} Merritt D, \& Ferrarese L.,
           2001b, MNRAS, 320, L30
   \bibitem[Nelson 2000]{ne00}Nelson, C. H., 2000, ApJ, 544, L91
   \bibitem[Nelson \& Whittle 1995]{nw95}Nelson, C. H., \& Whittle, M., 
           1995, ApJS, 99, 67
   \bibitem[Nelson \& Whittle 1996]{nw96}Nelson, C. H., \& Whittle, M., 
           1996, ApJ, 465, 96
   \bibitem[Osterbrock \& Pogge 1985]{op85}Osterbrock, D. E., \& Pogge, R. W.,
           1985, ApJ, 297, 166
   \bibitem[Ostriker 2000]{ost00}Ostriker, J. P., 2000, PRL, 84, 5258
   \bibitem[Puchnarewicz et al. 1992]{puch92}Puchnarewicz, E. M., et al.
           1992, MNRAS, 256, 589
   \bibitem[Puchnarewicz et al. 2001]{pun01} Puchnarewicz, E. M., Mason, K. 
           O., Siemiginowska, A., Fruscione, A., Comastri, A., Fiore, F.,
           \& Cagnoni, I., 2001, ApJ, 550, 644
   \bibitem[Sarzi et al. 2000]{sar00}Sarzi, M., et al. 2000, ApJ, 550, 65
   \bibitem[Silk \& Rees 1998]{sr98}Silk, J. \& Rees, M. J., 1998, A\&A,
           331, L1
   \bibitem[Ulvestad et al. 1995]{uag95}Ulvestad, J. S.,
           Antonucci, R. R. J., Goodrich R., 1995, AJ, 109, 81
   \bibitem[Unger et al. 1987]{u87}Unger, S. W., Lawrence, A., Wilson, 
           A. S., Elvis, M., Wright, A. E., 1987, MNRAS, 228, 521
   \bibitem[Veilleux 1991]{vei91} Veilleux, S., 1991, ApJS, 75, 383
   \bibitem[Veron-Cetty et al. 2001]{vvg} Veron-Cetty, M.-P., Veron, P., 
           \& Goncalves, A. C., 2001, A\&A, 372, 730
   \bibitem[Wandel 1999]{wan99}Wandel, A., 1999, ApJL, 519, L39
   \bibitem[Whittle 1985]{w85} Whittle, M., 1985, MNRAS, 216, 817
   \bibitem[Zheng et al. 1999]{zheng99}Zheng, Z., Wu, H., Mao, S., Xia, X. Y.,
            Deng, Z. G., \& Zou, Z. L., 1999, A\&A, 349, 735
\end{thebibliography}
\end{document}